\documentclass[pr,twocolumn,showpacs]{revtex4}
\usepackage{bm}
\usepackage{graphicx}
\usepackage{amsmath}
\usepackage{eufrak}
\usepackage{color}
\usepackage{upgreek}
\usepackage{subfigure}
\usepackage[unicode=true,colorlinks=true,citecolor=blue]{hyperref}
\usepackage{placeins}
\usepackage{lipsum}

\newcommand{\nix}[1]{}

\begin{document}

\title{Circular and linear magnetic quantum ratchet effects in dual-grating-gate
CdTe-based nanostructures}

\author{P.\,Faltermeier,$^1$  G.V. Budkin,$^2$  S.\,Hubmann,$^1$ 
V.V.\,Bel'kov,$^2$  L.E. Golub,$^2$ E.L.~Ivchenko,$^2$  Z.~Adamus,$^3$ G. Karczewski,$^3$ T. Wojtowicz,$^{3,4}$ D.A. Kozlov,$^{5}$ D.\,Weiss,$^1$ and S.D.\,Ganichev$^1$}

\affiliation{$^1$Terahertz Center, University of Regensburg, 93040 Regensburg, Germany}

\affiliation{$^2$Ioffe Institute, 194021 St.\,Petersburg, Russia}

\affiliation{$^3$ Institute of Physics, Polish Academy of Sciences, al. Lotnik\'{o}w 32/46, PL 02-668 Warszawa, Poland}

\affiliation{$^4$ International Research Centre MagTop, al. Lotnik\'{o}w 32/46, PL 02-668 Warszawa, Poland}

\affiliation{$^5$ A.V. Rzhanov Institute of Semiconductor Physics, 630090 Novosibirsk,  Russia}

\begin{abstract}
We report on the observation and systematic study of polarization sensitive magnetic quantum 
ratchet effects induced by alternating electric fields in the terahertz frequency range.
The effects are detected  in 
(Cd,Mn)Te-based quantum well (QW) structures 
with inter-digitated  dual-grating-gate (DGG) lateral superlattices. 
A  $dc$ electric current excited by \textit{cw} terahertz laser radiation
shows $1/B$-periodic oscillations with an amplitude much larger than the photocurrent at zero
magnetic field. Variation of gate voltages applied to individual grating gates of the 
DGG enables us to change  the degree and the sign of the lateral asymmetry in a controllable way. 
The data reveal that the photocurrent reflects the degree of lateral asymmetry induced by 
different gate potentials. 
We show that the magnetic ratchet photocurrent includes the Seebeck thermoratchet effect as 
well as the effects of 
``linear'' and ``circular'' ratchets, which are sensitive to the corresponding polarization of the 
driving electromagnetic force.
Theoretical analysis
performed in the framework of 
semiclassical approach and taking into account Landau quantization 
describes  the experimental results well.
\end{abstract}

\pacs{
 73.21.Fg,
 78.67.De,
 73.63.Hs
}
\maketitle

\section{Introduction}

Classical and quantum ratchets are spatially noncentrosymmetric systems 
which are able to transport classical or quantum particles in the absence of an average macroscopic force, for reviews see e.g.~Refs.\,\cite{prost97,reimann,hanggi,Denisovhanggi}. 
In semiconductors and semiconductor nanostructures, driving the system out of  thermal equilibrium, e.g. by high frequency alternating electric field, causes a $dc$ electric current. 
The required lack of inversion symmetry can be fulfilled
either by making use of an built-in asymmetry induced by the crystallographic
structure (in this case the ratchet effects are called photogalvanic effects) or
by artificial lateral superlattices superimposed on typically two-dimensional 
semiconductor materials. 
 Such superlattices have been realized in a great variety 
of forms which allow one to explore the origin of ratchet effects~\cite{buttiker,buttiker2,grifoni,kotthaus,samuelson,Chepel,Chepelianskii,Kvon,Olbrich_PRL_09,Popov2010,Olbrich_PRB_11,
Review_JETP_Lett,KannanPortal12,Nalitov,PopovAPL2013,Rozhansky2015,PopovIvchenko,ratchet_graphene,PRB2017} as well as to 
apply them to detect terahertz radiation~\cite{otsuji,det2,otsuji2,Popov_Otsuji_Knap}.  
The latter, besides high sensitivity and short response times, offers new functionalities being a good candidate for all-electric detection of the radiation polarization state including radiation 
helicity being so far realized applying photogalvanics in QW structures~\cite{ellipticitydetector,ellipticitydetector2} and HEMT structures~\cite{ellipticitydetector3,Otsuji_Ganichev}. 

Most recently it has been demonstrated that application of an external magnetic field to QW structures 
with DGG lateral superlattices results in  $1/B$-periodic oscillations of the electric ratchet  current with an amplitude much larger than the current at zero magnetic field~\cite{PRB2017,JETP_Lett_review}.  The terahertz radiation induced magnetic ratchet current reported in Ref.~\cite{PRB2017} has been shown to be caused by the Seebeck ratchet (or thermoratchet) effect, so far detected in zero magnetic field only~\cite{Review_JETP_Lett}. 
It stems from the combined effect of  radiation-induced inhomogeneous heating of the  two-dimensional electron gas and the grating-induced electrostatic electron potential $V(x)$, where $x$ is the in-plane coordinate in the direction  perpendicular to the DGG metallic stripes. 
The inhomogeneous heating
is caused by the near-field space modulation of the electric field ${\bm E}(x)$ of the radiation. The coordinate dependent field acts on the electron gas changing the local effective electron temperature to  $T(x) = T + \delta T(x)$, where $T$ is the equilibrium temperature. 
The inhomogeneous heating causes the diffusion of electrons from warmer to colder regions and forms a nonequilibrium electron density profile $\delta N(x)$. In short, the Seebeck ratchet current is a drift of the nonequilibrium correction $\delta N(x)$ in the electric field $- (1/e) dV(x)/dx$ of the space modulated electrostatic potential. 
The Seebeck ratchet current is insensitive to the radiation polarization state and, in the presence of quantizing magnetic field, follows the $1/B$-periodic  oscillations of the longitudinal resistance.

Here we report on the observation and study of  polarization sensitive magnetic quantum ratchet effects driven by linearly polarized radiation (linear magnetic ratchet effect) or/and by the radiation helicity (circular magnetic ratchet effect).
These effects are demonstrated for (Cd,Mn)Te/(Cd,Mg)Te diluted magnetic heterostructures  with superimposed lateral asymmetric superlattices. 
Similarly to the magnetic thermoratchet effect, applying a magnetic field $B$ along the growth direction we have observed that
the polarization sensitive magnetic ratchet currents exhibit sign-alternating $1/B$-periodic oscillations with amplitudes substantially larger than the ratchet signal at zero magnetic field.
 In contrast to the thermoratchet the mechanisms of these ratchet effects are unrelated to  electron-gas heating. 
They arise  from the phase shift between the periodic electrostatic potential and the periodic radiation  field resulting from near field diffraction. 
The ratchet currents are determined by the in-plane orientation of the electric field vector (linear magnetic ratchet effect) or by
the radiation helicity (circular magnetic ratchet effect). 
They appear because the carriers in the laterally modulated quantum wells 
move in both in-plane directions $x,y$ and are subjected to the action of the two-component electric field $\bm E_\omega =(E_{\omega,x}, E_{\omega,y})$.  Theoretical analysis, considering the
magnetic quantum ratchet effect in the framework of semiclassical
approach~\cite{JETP_Lett_review,PRB2017,Budkin_Golub}, describes   the experimental results well. It shows that 
the observed magneto-oscillations with enhanced photocurrent amplitude 
result from Landau quantization. Furthermore, for (Cd,Mn)Te at low
temperatures, the oscillations are affected by the exchange enhanced Zeeman splitting 
in diluted magnetic heterostructures. 

The paper is organized as follows. In Sec.~\ref{sample} we  describe the investigated samples, the experimental technique and results for zero magnetic field. In Sec.~\ref{magnetic} we discuss the results on  magnetic ratchet effects generated by linearly/circularly polarized THz radiation. In the following Sec.~\ref{theory} and Sec.~\ref{discussion} we present the theory 
 and compare its results with the experimental data. Finally,  we summarize the results and give an outlook for future studies of the ratchet effects (Sec. \ref{summary}).

\section{Samples, experimental technique and photocurrents at zero magnetic field}
\label{sample}

\subsection{Quantum well structures}

\begin{figure}[t]
\includegraphics[width=\linewidth]{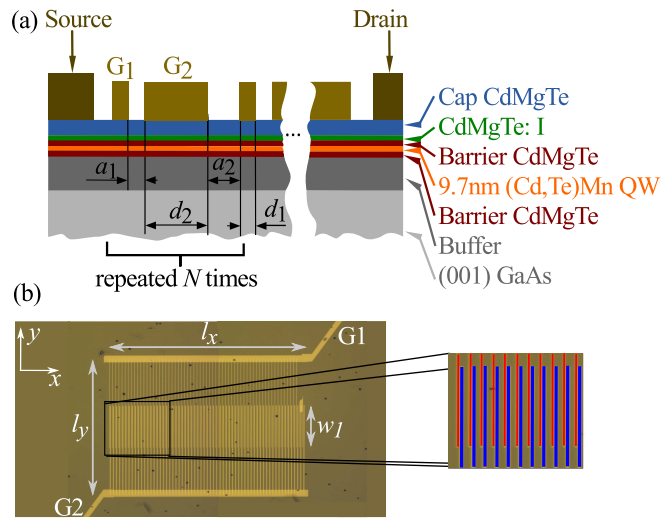}
\caption{ (a) Cross-section schematic view of CdMgTe/ (Cd,Mn)Te/CdMgTe QW with a 
dual-grating gate superlattice formed by metal stripes deposited
on top of the QW structure. (b) Photograph of the sample together with schematic top view of the dual-grating gate
superlattice. Narrow red and wide blue lines sketch the 
interconnected top gates having different widths $d_1$ and $d_2$.}
\label{Fig_1_structure}
\end{figure}

\begin{figure}[h]
\includegraphics[width=\linewidth]{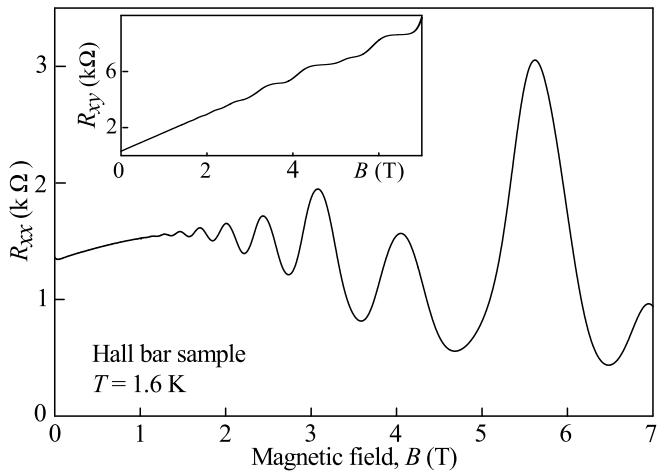}
\caption{ Longitudinal resistance $R_{xx}$
measured vs. magnetic field at T =1.6 K in the (Cd,Mn)Te QW unpatterned Hall bar sample. Inset shows Hall resistance $R_{xy}$ }
\label{Fig_0_transport}
\end{figure}

Ratchet currents have been studied in (Cd,Mn)Te/CdMgTe QW structures with doubly 
inter-digitated grating gates (DGG)  G$_1$ and G$_2$.
The cross-section of the structure, a sketch and a microphotograph of DGG
are shown in Fig.~\ref{Fig_1_structure}. Structures have been grown by molecular 
beam epitaxy on (001)-oriented GaAs 
substrates~\cite{Crooker,Egues,Jaroszynski2002,DMSPRL09}.  We used a   QW of 9.7\,nm width with Cd$_{0.76}$Mg$_{0.24}$Te alloy 
as a barrier material.  The samples are modulation doped with Iodine 
donors introduced into the top barrier at 10\,nm distance from the QW.  
Doped region has been overgrown by a 50~nm Cd$_{0.76}$Mg$_{0.24}$Te  undoped cap layer.
Two evenly spaced Cd$_{1-x}$Mn$_x$Te thin layers   were inserted
during the  QW growth applying the digital alloy technique\,\cite{Kneip2006}.
Incorporation  of Mn atoms 
into the QW region leads to a strong enhancement of the
effective $g$-factor of band carriers, and hence to an enhanced Zeeman splitting. 
Analysis of the magneto-photoluminescence data~\cite{DMSPRL09} obtained on the 
samples made from the same wafer as used in the current study
have shown that the spin splitting is well described by 
the modified Brillouin function~\cite{Gaj79,Fur88}
by using the effective average concentration of Mn
in the digital alloy $\bar{x}$ = 0.015.
The samples have also been characterized by electrical transport
measurements. At low temperatures pronounced Shubnikov-de-Haas (SdH) oscillations 
and  well resolved quantum Hall plateaus have been observed, see Fig.~\ref{Fig_0_transport}. 
The density of the two-dimensional electron gas, $n_e$,  
and the electron mobility, $\mu$, determined at liquid helium
temperature (4.2 K) are $n_e = 6.6 \times 10^{11}$~cm$^{-2}$
and $\mu = 9.5 \times 10^3$~cm$^{2}$/Vs. More details on  structures 
characteristics can be found in Ref.~\cite{PRB2017}.

\subsection{Inter-digitated  dual-grating-gate structures}

The DGG are formed by electron beam lithography, followed by deposition of 25\,nm-thick gold films and subsequent lift-off.
The grating-gate supercell consists of two metal stripes having 
widths $d_1 = 1.85~\mu$m and  $d_2 = 3.7~\mu$m, see Fig.~\ref{Fig_1_structure}. The corresponding spacings in sample \#A (sample \#B)  are $a_1 = 2.8~\mu$m ($3.45~\mu$m) and
$a_2 = 5.65~\mu$m ($5.0~\mu$m). 
The supercell is  repeated $65$ times to produce a lateral asymmetric superlattice with the period $d = d_1 + a_1 +d_2 +a_2 = 14~\mu$m, see Refs.~\cite{Olbrich_PRB_11,ratchet_graphene,staab2015}.
All narrow (top gate G1) and wide  (top gate G2) grating stripes have been connected by  additional gold bars [wide horizontal lines in Fig.~\ref{Fig_1_structure}(b)]
so that wide and narrow gate stripes can be biased independently, see inset in
Fig.\,\ref{Fig_2_alpha_sA}. The distance between the
connecting bars is $l_y = 575 \,\mu$m and the length 
$l_x = 905\,\mu$m. The width of the area with both gates present is $w_1 = 190\, \mu$m, see Fig.~\ref{Fig_1_structure}(b).
The size of the 
complete DGG structure is about $905 \times 190 \,\mu$m$^2$.

Two  pairs of ohmic contacts have been prepared to measure the photocurrent perpendicularly  to the metal stripes ($J_x$, $x$-direction) and parallel to them ($J_y$, $y$-direction), see inset in Fig.\,\ref{Fig_2_alpha_sA}.  Same contact pairs  have been used for magneto-transport measurements  accompanying the measurement of magnetic field induced photocurrent. 
The samples were placed into an optical cryostat with $z$-cut
crystal quartz windows and superconducting magnet. The
magnetic field ${\bm B}$ up to 7~T could be applied  normal to  the QW plane.

Magneto-transport measurements show that the application of gate voltages to individual gates does 
not visibly influence  the period of $1/B$-periodic oscillations of the longitudinal resistance, $R_{xx}$.
This is due to the fact that the area of DGG fingers is a very small fraction of the whole sample
area. Note, that  while the overall transport characteristics obtained at different cooldowns is the same
the value of the effective gate voltages acting on the electrons in 2DEG may change by $\pm 0.1$~V. 

\subsection{Methods}

The photocurrents are  generated  in samples cooled down to liquid helium temperature applying  THz radiation of a  continuous wave (\textit{cw}) molecular optically pumped laser~\cite{Ganichev93,Kvon2008,Kohda2012} operating at frequency ${f=2.54}$~THz (photon energy $\hbar\omega = 10.4$~meV,  wavelength ${\lambda =118~\mu}$m). 
The radiation photon energy is  smaller than the
band gap as well as the size-quantized subband separation. Thus,
the radiation induces only indirect (Drude-like) optical transitions in the lowest
conduction subband (Drude-like free carrier absorption).
The measurements of magnetic-field-induced photocurrents are carried
out under excitation of the (001)-grown QW samples with
polarized terahertz radiation at \textit{normal} incidence. This experimental geometry is chosen to exclude any effects in the area outside the DGG structure  known to cause photocurrents at zero magnetic field and for magnetic field normal to the QW plane~\cite{Review_JETP_Lett,PRB2017}. Measurements applying a pyroelectric camera~\cite{Ganichev1999,Schneider04} have shown that the spatial beam distribution has an almost Gaussian profile
with a beam spot on the sample of about $1.3$~mm diameter. 

The incident power about 30~mW is modulated at about 75~Hz  by an optical chopper. Taking into account sizes of the superlattice and the beam spot we obtain that the power irradiating the structure  is  $P \approx 4$\,mW. 
The photoresponse is measured in \textit{unbiased} samples by the voltage drop $U$ across a load resistor $R_L = 50$~Ohm~ using standard lock-in technique. 
The benefit of using of the small value of $R_L\ll R_s$, where $R_s$ is the sample resistance, is that the detected signal is unaffected by the sample resistance variation and is just proportional to the electric current generated by the THz radiation.  The current is calculated via $J = U/R_L$.

The laser radiation is initially linearly polarized along the $x$-axis. To explore the polarization dependence of
the ratchet effect we used crystal quartz  $\lambda/4$ plate as well as one-dimensional metal mesh polarizers~\cite{book}.
To vary the azimuthal angle $\alpha$ of linearly polarized radiation we rotated the mesh polarizer placed behind 
the $\lambda/4$ plate providing a circularly polarized radiation.
The  angle $\alpha$ is defined as an angle between radiation electric field vector and direction $x$, which is perpendicular to the metal stripes.
To study the helicity dependence of the signal the $\lambda/4$ plate was rotated by an angle $\varphi$ between the laser polarization plane and the optical axis of the plate. 
In this geometry, the radiation helicity is 
varied as $P_{\rm c0} \propto \sin{2 \varphi}$. 
Note that for $\alpha=0$ and as well  for $\varphi=0$ radiation is linearly polarized.

\subsection{Photocurrent at zero magnetic field}
\label{zero}

\begin{figure}
    \includegraphics[width=\linewidth]{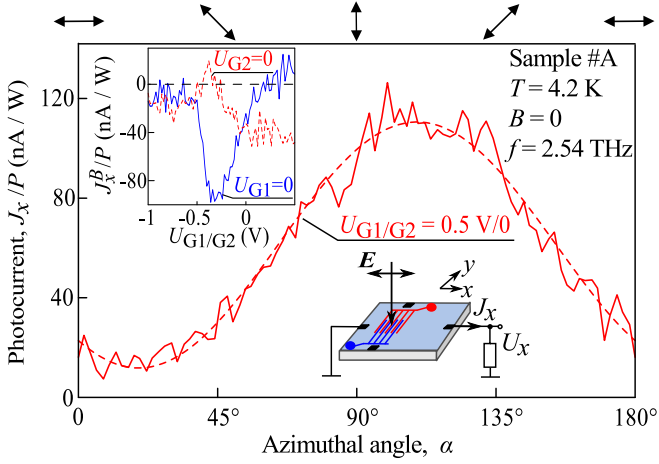}
    \caption{
Photocurrent $J_x$ normalized to the radiation power $P$ measured in sample~\#A at zero magnetic field as a function of the azimuthal angle $\alpha$.  
Arrows on top illustrate the polarization plane orientation for several azimuthal angles $\alpha$. The data are obtained for gate voltages $U_{\rm G1} = 0.5$~V (narrow gate stripes) and $U_{\rm G2} = 0$ (wide gate stripes). For briefness from here on we will indicate the magnitudes of the gate voltages applied to sublattices as following $U_{{\rm G1}/{\rm G2}} = 0.5\,$V$/ \,0$. Dashed curve is 
 the result of fitting the Eq.~(\ref{FITalpha}) to experimental data. The inset in the left upper corner shows the dependence of the magnitudes $J^B_x$ on the gate voltage $U_{\rm G1}$ ($U_{\rm G2}$) obtained for zero voltage applied to the other subgate $U_{\rm G2}$ ($U_{\rm G1}$). The bottom inset sketches the  experimental geometry. }
\label{Fig_2_alpha_sA}
    \end{figure}

\begin{figure}[h!]
    \includegraphics[width=\linewidth]{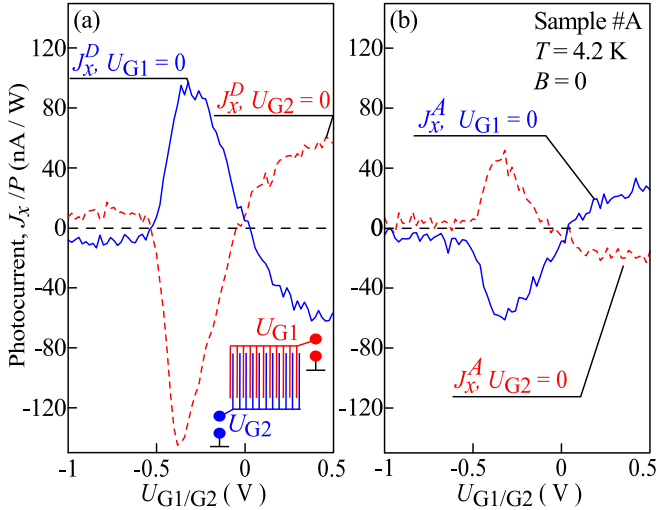}
\caption{
 The normalized photocurrent contributions $J^D_x$ and $J^A_x$ 
	measured in sample~\#A at zero magnetic field. Magnitudes of $J^D_x$ and $J^A_x$ are
  normalized to the radiation power  $P$ and plotted against the gate voltage $U_{\rm G1}$ ($U_{\rm G2}$) 
for zero voltage applied to the other subgate $U_{\rm G2}$ ($U_{\rm G1}$). Inset in the panel (a) shows schematically the gate structure with gate potentials applied to the narrow stripes 
(gate ${\rm G1}$) and the wide stripes (gate ${\rm G2}$).	
}
    \label{Fig_3_gate_sA}
\end{figure}

\begin{figure}
    \includegraphics[width=\linewidth]{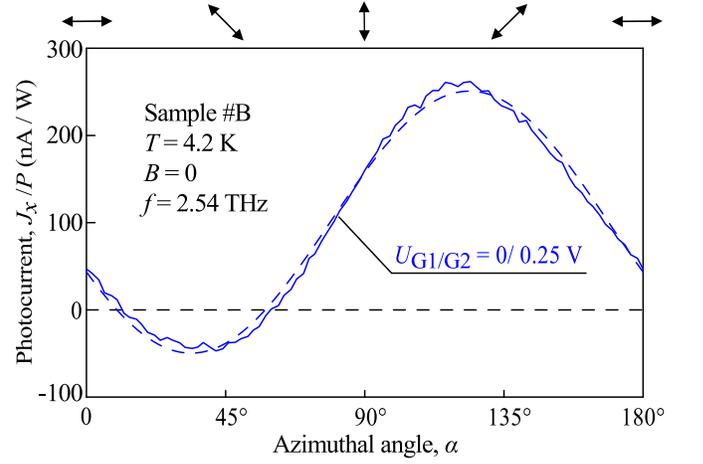}
    \caption{
	Photocurrent $J_x$ normalized to the radiation power $P$ measured in sample~\#B at zero magnetic field as a function of the azimuthal angle $\alpha$.   The data are obtained for gate voltages $U_{{\rm G1}/{\rm G2}} = 0\,/ \,0.25$\,V.  Dashed curve is the result of fitting the Eq.~(\ref{FITalpha}) to experimental data. Arrows on top illustrate the polarization plane orientation for several azimuthal angles $\alpha$.}
\label{Fig_4_alpha_sB}
    \end{figure}

We  start with the description of the results obtained for sample~\#A excited by  linearly polarized radiation at zero magnetic field. Illuminating the DGG  
by normally incident radiation we detect the photocurrent $J_x$ and $J_y$
measured in directions normal and parallel to the DGG stripes, respectively.
 The current vanishes for a beam shifted to the unpatterned area. This is in agreement with the discussion above underlining 
that  due to symmetry arguments photocurrents can not be generated in unpatterned structures, as was also observed in previous results obtained on the same (001)-grown (Cd,Mn)Te/CdMgTe QW structures
(sample~\#A in Ref.~\cite{PRB2017}). The overall behavior of both $J_x$ and $J_y$ photocurrents is very similar
but the magnitude of the photocurrent in $y$-direction is about 2 to 10 times smaller as that in the $x$-direction. Therefore, for this sample we present the data for $J_x$, only. As shown in Fig.~\ref{Fig_2_alpha_sA}, the current
consists of the polarization independent contribution and 
contributions varying upon rotation of the radiation 
electric field vector defined by the azimuthal angle $\alpha$. The overall polarization dependence can be well fitted by 
 \begin{equation} \label{FITalpha}
J_{x,y}(\alpha) =  J^A_{x,y} \sin{2\alpha} + J^B_{x,y} \cos{2\alpha} + J^D_{x,y}\:,
 \end{equation}
 where $J^A_{x,y}$, $J^B_{x,y}$, and $J^D_{x,y}$ are fitting parameters.
We note, that $\cos{2\alpha}$ and $\sin{2\alpha}$ represent the  Stokes 
parameters and describe the degree of linear polarization in the coordinate axes $x,y$ 
and within the system rotated about an angle of 45$^\circ$, respectively~\cite{Stokes1,BelkovJPCM}.
The fact that the photocurrent is detected at normal incidence of patterned structure  indicates that the photocurrent is caused by the ratchet effect.

An ultimate proof for ratchet effect as a cause of the photocurrent, however, comes from the 
experiments applying different gate voltages to the DGG subgates. 
The ratchet effects are obviously expected to be strongly dependent
on the in-plane asymmetry of the electrostatic potential
being proportional to the averaged product~\cite{Review_JETP_Lett}, 
 \begin{equation} \label{Xi}
 \Xi = \overline{\frac{dV}{dx} |\bm E(x)|^2} 
\end{equation}
 of the derivative of the coordinate dependent electrostatic potential $V(x)$ and the 
distribution of the electric near-field  $\bm E(x)$~\cite{Review_JETP_Lett,Nalitov,PopovAPL2013,PopovIvchenko}. 
In the following, we call the  parameter $\Xi$ the \emph{lateral asymmetry}. 
The value of $\Xi$  may change sign depending on electrostatic potential $V(x)$. Consequently a variation of 
individual gate voltages should result in a change of the ratchet current including 
reversal of its direction. Sweeping one subgate bias voltage, e.g. $U_{\rm G1}$,  from negative to positive
 and holding another one at zero should cause the 
sign inversion of the ratchet current. 
Exactly  this behavior has been observed in the experiment for all three
contributions, see the inset in Fig.~\ref{Fig_2_alpha_sA} and Fig.~\ref{Fig_3_gate_sA}.
Furthermore, reversing the circuit in the way that now $U_{\rm G2}$
is swept and $U_{\rm G1} = 0$ causes the reversal of the photocurrent direction, see~Fig.~\ref{Fig_2_alpha_sA}.  Note that for unbiased gates a non-zero built-in electrostatic potential is formed due to the presence of metal stripes in the vicinity of QW.
 
We also note that proportionality of the observed current to 
the lateral asymmetry parameter $\Xi$ demonstrates that a possible
contribution due to  edge photogalvanic effects, similar to the polarization-sensitive photocurrents in graphene flakes~\cite{Karch_PRL_2011,Glazov2014},
plays no essential role in the described experiments. 

Figures~\ref{Fig_2_alpha_sA} and ~\ref{Fig_3_gate_sA} reveal that in sample \#A  the dominating contribution to the total photocurrent  comes from the polarization independent current $J^D_x$, being previously shown to be caused by the Seebeck ratchet~\cite{PRB2017}.  Nevertheless polarization dependent photocurrent contributions $J^A$ and $J^B$ are clearly detected in this sample for almost all combinations of the gate voltage. First of all we emphasize that the polarization dependence can not be simply attributed to the Seebeck ratchet effect changing its magnitude due to the change of the radiation reflection from the DGG. The latter would results in a signal maximum for the radiation electric field aligned perpendicularly to the grating's stripes ($\alpha = 0$), whereas in experiment we observed just  the opposite behavior: the maximum signal is achieved for electric field $\bm E_0$ parallel to the stripes in $y$-direction ($\alpha = 90^\circ$), see Fig.~\ref{Fig_2_alpha_sA}.  
A more direct prove for the existence of the polarization dependent photocurrents is detected in sample \#B with the larger gates spacing $a_1$.
In this sample for a certain combination of gate voltages we detected change of the photocurrent sign upon variation 
of the azimuthal angle: the fact clearly demonstrating contribution of an origin different from the Seebeck ratchet, see Fig.~\ref{Fig_4_alpha_sB}.  While polarization dependent photocurrents ($J^A$ and $J^B$) for  
CdTe-based QWs at zero magnetic field are reported here for the 
first time  they have been detected  and discussed in details
for other materials (GaAs QWs, InAlAs/InGaAs/InAlAs/InP HEMT and graphene) in~Refs.~\cite{Olbrich_PRL_09,Olbrich_PRB_11,Review_JETP_Lett,Nalitov,ratchet_graphene}. 
 The present work is focused on  the polarization dependent \textit{magnetic} ratchet currents, 
therefore, a deeper analysis of the results for zero magnetic field is out of its scope.

\begin{figure}[h]
        \includegraphics[width=\linewidth]{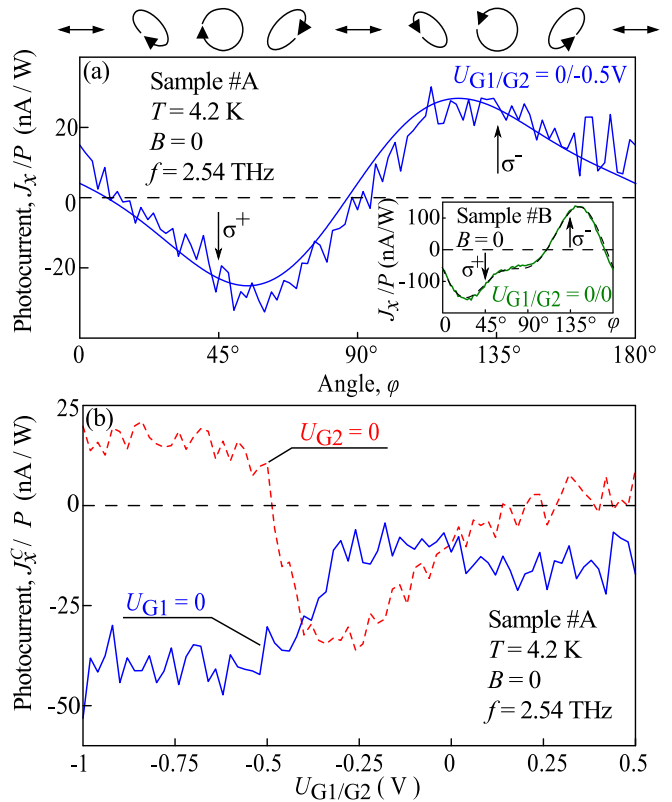}
        \caption{				
Photocurrent $J_x$ normalized to the radiation power $P$ measured in (Cd,Mn)Te QW samples for zero magnetic field. (a)  Photocurrent induced in sample~\#A as a function of the phase angle $\varphi$. The ellipses on top illustrate the polarization states for several values of $\varphi$. Vertical arrows indicate angles $\varphi$ for the right-handed ($\sigma^+$) and left-handed ($\sigma^-$) circularly polarized radiation. The data are obtained for gate voltages applied to narrow/wide gate stripes $U_{{\rm G1}/{\rm G2}} = 0/-0.5~V$.  Solid line is 
the result of fitting the Eq.~(\ref{circ_B0_fit}) to experimental data. Inset shows as well 
the polarization dependence of
the photocurrent detected in  sample~\#B  with both gates set  to zero voltage. Dashed line is 
the result of fitting the Eq.~(\ref{circ_B0_fit}) to experimental data. 
(b) Dependence of $J^C_x$ on the gate voltage $U_{\rm G1}$ ($U_{\rm G2}$) obtained at zero gate voltage $U_{\rm G2}$ ($U_{\rm G1}$) for sample~\#A.} 
  \label{Fig_5_phi_sAB}
    \end{figure}

Previous studies of ratchet effects in other materials (for review see Ref.~\cite{Review_JETP_Lett}) demonstrated that beside the photocurrents addressed above a contribution proportional to the radiation helicity may also exist. This contribution we also detected for (Cd,Mn)Te/CdMgTe QW samples.
Figure~\ref{Fig_5_phi_sAB}(a) shows photocurrent $J_{x}$ as a function
of the phase angle $\varphi$.  This dependence is measured for subgate voltages $U_{\rm G1} = 0$, $U_{\rm G2}=-0.5$~V for which  
contributions proportional to coefficients $J^A_x$, $J^B_x$, and $J^D_x$
are strongly suppressed, see the inset in Fig.~\ref{Fig_2_alpha_sA} and Fig.~\ref{Fig_3_gate_sA}. The data can be well fitted by 
\begin{eqnarray}\label{circ_B0_fit}
J_{x,y}(\varphi) &=& J^A_{x,y} \frac{\sin{4\varphi}}{2} + J^B_{x,y} \frac{1+\cos{4\varphi}}{2} \\&& + \ J^C_{x,y}\sin{2\varphi} + J^D_{x,y}\:, \nonumber
\end{eqnarray}
 where $J^C_{x,y}$ is a fitting parameter for the contribution proportional to the degree of circular polarization  $P_{\rm c0}$, i.e. the fourth Stokes parameter~\cite{Stokes1,BelkovJPCM}. Sweeping either $U_{\rm G1}$ or $U_{\rm G2}$
gate voltages we have proved that  helicity driven photocurrent is as well proportional to the degree of the electrostatic potential asymmetry.
This is shown in  Fig.~\ref{Fig_5_phi_sAB}(b). Note that in contrast to the currents obtained for linearly polarized light the gate dependencies
for $U_{\rm G1}=0$, $U_{\rm G2}\neq 0$ and $U_{\rm G2}=0$, $U_{\rm G1}\neq 0$ are not symmetric with respect to the abscissa. We also note that, like the polarization independent contribution, the curves $J^C_x(U_{\rm G1})$ and $J^C_x(U_{\rm G2})$ cross each other not only near zero-bias values but also for rather high negative voltages. 

Similar results have been obtained for sample~\#B. An example of the helicity dependence measured in this sample for $U_{\rm G1}=0$ or $U_{\rm G2}=0$ is shown in the inset in Fig.~\ref{Fig_5_phi_sAB}(a). It shows that  in this sample the data are also well fitted by Eq.~(\ref{circ_B0_fit}).  
It should be noted that the ratchet current can be detected even for  gate voltages applied to
both gates. As addressed above for unbiased gates, the required lateral asymmetry  comes from a non-zero
built-in electrostatic potential which is formed due to the presence of   metal stripes in the QW vicinity. This potential is shown to be particularly large in sample~\#B.

To summarize this part, the measurements for zero magnetic field show that excitation of DGG CdTe-based superlattices
by  THz radiation results in  polarization independent and three polarization dependent $dc$ ratchet photocurrents which can efficiently be controlled by voltages applied to individual subgates. The overall  behavior of the photocurrents is in qualitative agreement with that of the electronic ratchet effects observed in semiconductor QW structures and graphene with a lateral superlattice and these ratchet effects can well described by the microscopic theory
developed in Refs.~\cite{Olbrich_PRL_09,Olbrich_PRB_11,Review_JETP_Lett,ratchet_graphene}.

\section{Magnetic field induced photocurrent} \label{magnetic}

\begin{figure}[ht]
     \includegraphics[width=\linewidth]{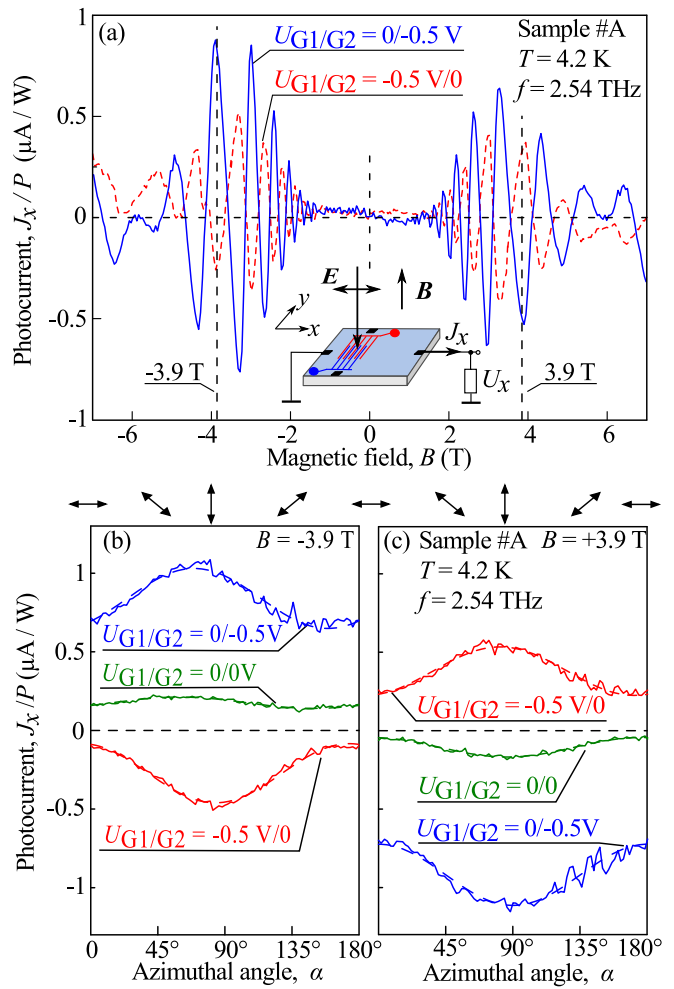}
     \caption{
		Panel (a): Normalized photocurrent $J_x/P$ as a function of the magnetic field $B$ measured in sample~\#A. The data are obtained for linearly polarized radiation with electric field $\bm E$ oriented perpendicularly to the grating stripes.
Solid blue and dashed red lines show the data for the gate voltages  
$U_{\rm G1/G2} = 0/-0.5$~V and $U_{\rm G1/G2} = -0.5$~V/0, respectively.  
 The inset shows the experimental setup. 				
Panels (b) and (c): Dependencies of the magneto-photocurrent $J_x$  on the azimuthal angle $\alpha$ measured in sample~\#A.   The data are obtained for two magnetic field strengths, $B = \pm 3.9$~T, corresponding to oscillation peaks in magnetic field dependencies.
The dashed curves are 
the result of fitting the Eq.~(\ref{FITalpha}) to experimental data obtained for different lateral asymmetries of the electrostatic potential. Arrows on top illustrate the polarization plane orientation for several $\alpha$.	}
				\label{Fig_6_alpha_sA}
    \end{figure}

Now we turn to the main part of the paper devoted to polarization dependent magnetic quantum ratchet effects.  
Applying magnetic field normal to  the QW plane we observed that the photocurrent increases drastically  
and, at high magnetic fields, exhibits sign-alternating $1/B$-periodic oscillations. 
This is shown in Fig.~\ref{Fig_6_alpha_sA}(a)
for linearly polarized radiation with electric field vector oriented normal to the DGG stripes. 
This figure also reveals that reversal of  
lateral asymmetry of the electrostatic potential causes the sign change of the photocurrent oscillations.
 Note that the positions of oscillation 
maxima/minima  remain unchanged for all potentials used in our study of the sample. 

			\begin{figure}[ht]
        \includegraphics[width=\linewidth]{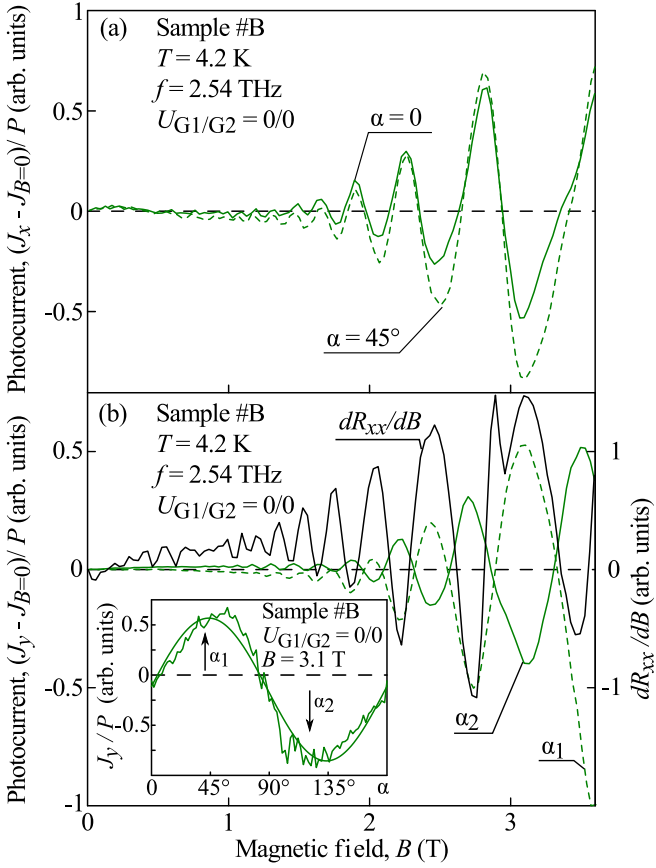}
\caption{ 
Magneto-oscillations of the  normalized photocurrents $J_x/P$ [panel (a)] and $J_y/P$ [panel (b)] measured in sample~\#B for several azimuthal angles $\alpha$ at which photocurrents approach maximum/minimum values. The data are obtained for zero voltages applied to both gates. Black solid line in panel (b) shows the first derivative of 
longitudinal resistance, $dR_{xx}/d B$, measured in the van-der-Pauw geometry. The inset in panel (b) shows the normalized photocurrent $J_y/P$ as a function of the azimuthal angle $\alpha$ measured for magnetic field $B = 3.1$~T which corresponds to a peak of magneto-oscillations. Solid curve is 
the result of fitting the Eq.~(\ref{FITalpha}) to experimental data.}
        \label{Fig_7_bdep_sB}
    \end{figure}	
		
Rotation of the electric field orientation results in a variation of the photocurrent magnitude. This is shown in Figs.~\ref{Fig_6_alpha_sA}(b) and (c) for two magnetic field strengths corresponding to peaks of magneto-oscillations [see vertical dashed lines in panel (a)]. 
The overall polarization dependence of the photocurrent can be well fitted by Eq.~(\ref{FITalpha}) with fitting parameters  $J^A_x, J^B_x$, and  $J^D_x$ which now depend on the magnitude and sign of the external magnetic field. Figures~\ref{Fig_6_alpha_sA}(b) and (c) show that in sample~\#A the polarization independent photocurrent $J^D_x$ provides a dominant contribution. This photocurrent has been studied and discussed in details in Ref.~\cite{PRB2017}. It has been shown to be caused by the Seebeck magnetic quantum ratchet effect and will be not discussed here. Our present study is focused on polarization dependent parts.  

Figures~\ref{Fig_6_alpha_sA}(b) and (c) reveal that, similarly to
the polarization independent Seebeck magnetic quantum ratchet effect, 
polarization dependent parts change the sign upon reversal of the lateral electrostatic potential, see the curves for $U_{\rm G1/G2} = -0.5$~V/0 and $0 /-0.5$~V. This observation provides an evidence that the polarization dependent part of the photocurrent 
is also driven by the ratchet effect. Due to a large contribution of the $J^D_x$ 
detected for sample~\#A the variation of the electric field orientation causes 
only the modulation of the photocurrent magnitude. 

In contrast to sample~\#A, in sample~\#B with the larger spacing $a_1$ we have observed that 
relative contribution of $J_D$ is substantially diminished for the photocurrent 
measured in the $y$-direction. Moreover, the magnitudes of the contributions $J_x$ and $J_y$ 
are comparable in this sample. In  the direction perpendicular to the DGG stripes of sample~\#B
photocurrent is dominated by the polarization independent magnetic Seebeck ratchet effect and 
the variation of the angle $\alpha$ only slightly affects the oscillation 
magnitudes without changing their sign, see Fig.~\ref{Fig_7_bdep_sB}(a).
For the $y$-direction, however,
changing of the angle $\alpha$ results in the sign change of the photocurrent $J_y$
revealing a dominating contribution of the polarization dependent photocurrent, 
see Fig.~\ref{Fig_7_bdep_sB}(b) and inset in this panel. Here and  in the remaining plots  
a small contribution of the zero magnetic field photocurrent, $J_{B=0}$, is subtracted  
from the total current. Polarization dependence of $J_y$ 
obtained for one of the peaks of magneto-oscillations [see vertical dashed line in panel (b)] demonstrates that 
in the structure~\#B with zero applied voltages  $U_{\rm G1}=U_{\rm G2}=0$
the polarization independent current is almost negligible 
and the photocurrent is dominated by 
the contribution proportional to the coefficient
$J^A_y$. Figure~\ref{Fig_7_bdep_sB}(b) also reveals that 
magneto-oscillations measured for angles $\alpha_1$ and $\alpha_2$ indicated by vertical arrows in the inset have opposite signs for all peaks. 
		
At last but not at least we discuss the helicity driven magnetic photocurrents. 
While the influence of the radiation helicity on magnetic
quantum ratchets has been detected in both samples, similarly to the linear magnetic quantum ratchet, 
it is most pronounced in sample~\#B and the $y$-direction. 
Therefore,  here we focus on these data, and the results on sample~\#A are presented in the Appendix~\ref{appendix}.  
Figure~\ref{Fig_8_bdep_sB}(a) shows magnetic field dependencies of the photocurrents measured 
for right- and left-handed polarized radiation in directions parallel to the DGG stripes. 
The figure reveals that magneto-oscillations have opposite sign for opposite helicities. The overall
polarization dependence obtained by variation of the phase angle $\varphi$ can be well fitted by Eq.~(\ref{circ_B0_fit})
with magnetic field dependent coefficients $J^A, J^B, J^C, J^D$, see inset in Fig.~\ref{Fig_8_bdep_sB}(b). For purely right and left handed circular polarization, the terms proportional to $J^A_{x,y}$ and $J^B_{x,y}$ in Eq.~\eqref{circ_B0_fit} vanish and the current $J^D_{x,y}$ is independent of the helicity so that  
subtraction of these photocurrents yields 
$J^C_{x,y} = [J_{x,y}(\sigma^+) - J_{x,y}(\sigma^-)]/2$. The magnetic field dependencies of the circular photocurrent measured is shown in Fig.~\ref{Fig_8_bdep_sB}(b).

 \begin{figure}[ht]
        \includegraphics[width=\linewidth]{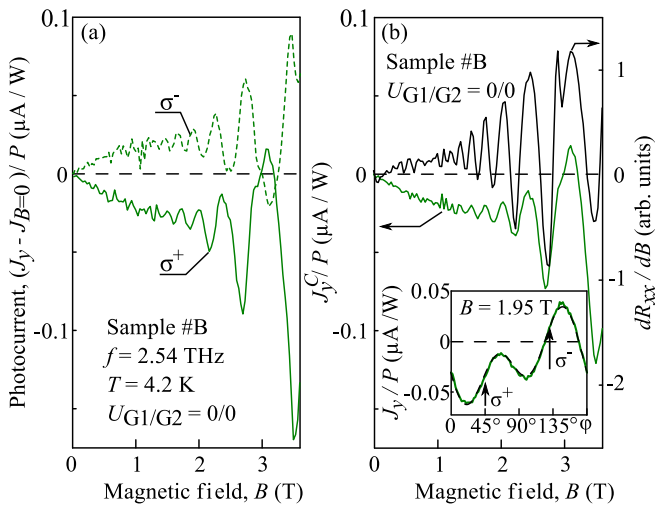}
        \caption{ 
				Magnetic field dependence of the normalized photocurrent $J/P$ measured in sample~\#B for circularly polarized radiation and zero gate voltages.
Panel (a) shows the photocurrent $J_y$ obtained for the
right- and left- circularly polarized radiation. 
Panel (b) shows circular photocurrent 
$J^C_{y} = [J_{y}(\sigma^+) - J_{y}(\sigma^-)]/2$. Black solid  line in the panel (b) shows the first derivative of 
longitudinal resistance, $dR_{xx}/d B$, measured in the van-der-Pauw geometry. The inset in panel (b) shows normalized photocurrents $J_y/P$ as a function of the phase angle $\varphi$. 
The data are obtained for magnetic field $B = 1.95$~T. 
Dashed line is  
the result of fitting the Eq.~(\ref{circ_B0_fit}) to experimental data.}
        \label{Fig_8_bdep_sB}
    \end{figure}

To summarize, experiments provide a consistent picture demonstrating a substantial role of the ratchet and magnetic ratchet effects driven by linearly and circularly polarized radiation. They are (i)~generated duef to the lateral asymmetry, (ii)~characterized by specific polarization dependencies for directions along 
and perpendicular to the metal stripes, and (iii)~changing their direction upon reversing the lateral asymmetry.
These results are in agreement with the theory of ratchet effects excited by the polarized THz electric field in lateral asymmetric  superlattices,  discussed in the next section.

\FloatBarrier
\section{Theory}
\label{theory}

The linear and circular magneto-ratchet effects induced 
by radiation of terahertz frequencies can be described in the framework of 
semiclassical approach and taking into account Landau quantization. 
Because the plasmonic resonance frequencies are well below the radiation 
frequency of 2.54~THz \cite{PRB2017}, we do not consider plasmonic effects. 
Below we present two independent microscopic mechanisms of ratchet effects dependent on the polarization of the incident radiation.

\subsection{Dynamic Carrier-Density Redistribution (DCDR)}

This first mechanism of the magnetic ratchet current generation is described within the formalism previously proposed for ratchet effects at zero magnetic field~\cite{Review_JETP_Lett,Nalitov}. 
It is related to the time-dependent, spatially periodic oscillation of the electron density 
\begin{equation} \label{oscillation}
\delta N(x,t) = \delta N_\omega(x) {\rm e}^{-{\rm i} \omega t} + {\rm c.c.} 
\end{equation}
which is  linear in both the radiation electric field ${\bm E}(x)$ and the lateral force $-dV (x)/dx$. The DCDR photocurrent is determined by the following working formula~\cite{Review_JETP_Lett,Nalitov}
\begin{equation}
\label{current}
	j_\alpha = {\partial \sigma^{0}_{\alpha \beta}\over \partial N} \overline{\delta N_\omega(x) E_{\omega, \beta}^*(x)} + {\rm c.c.}\:,
\end{equation}
where $\sigma^{0}_{\alpha \beta}$ is the static conductivity tensor $\sigma_{\alpha \beta}(\omega=0)$, and the overline implies an average over the spatial coordinate $x$. 

The  presence of the {\it static}  photoconductivity in Eq.~(\ref{current}) is a crucial point and requires a justification.
We can provide evidence to this according to the following general argument. 
Let us start with the linear nonlocal-in-time electric-current response 
\begin{equation}
\label{current2}
	j_\alpha (x,t) = \int\limits_{-\infty}^t dt' \sigma_{\alpha \beta}[t-t', N(x,t')] E_{\beta}(x,t') \:,
\end{equation}
where ${\bm \sigma}[t-t', N(x,t')]$ is the transient conductivity under a pulsed field for the instant electron density distribution $N(x,t')$. Presenting the latter as a sum of the equilibrium distribution $N_0(x)$ and the correction $\delta N(x,t)$ we obtain for the DCDR current
\begin{equation}
\label{current3}
	 j_\alpha (x,t) = \int\limits_{-\infty}^t dt' \frac{\partial \sigma_{\alpha \beta}(t-t', N)}{\partial N} \delta N(x,t') E_{\beta}(x,t') \:,
\end{equation}
where $N$ is the average electron density $\overline{N_0(x)}$.
Retaining in the product $\delta N(x,t') E_{\beta}(x,t')$ a time-independent part 
\[
\delta N_\omega(x) E^*_{\omega, \beta}(x) + \delta N^*_\omega(x) E_{\omega, \beta}(x)
\]
and taking into account that
\begin{eqnarray}
&&~~	\int\limits_{-\infty}^t dt' \frac{\partial \sigma_{\alpha \beta}(t-t', N)}{\partial N}\nonumber \\
= && \frac{\partial}{\partial N} \int\limits_{-\infty}^t dt' \sigma_{\alpha \beta}(t-t', N) 
\equiv  \frac{\partial \sigma^0_{\alpha \beta}(N)}{\partial N} \:, \nonumber
\end{eqnarray}
we finally get Eq.~(\ref{current}). In the following we 
neglect the energy dependence of the momentum scattering time $\tau$ \cite{footnotetau} and apply Eq.~(\ref{current}) for the degenerate two-dimensional electron gas subjected to the perpendicular magnetic field.

Given Eq.~(\ref{current}), the subsequent calculations are obvious
and allow us to derive equations describing linear as well as circular magnetic ratchet effects. Indeed the amplitude of the dynamic electron density $\delta N_\omega(x)$ is found from the continuity equation 
\begin{equation}
	- i \omega e \delta N_\omega(x) + {\partial \over \partial x} j_{\omega,x}(x)=0
\end{equation}
and Ohm's law for the {\it ac} current
\[
 j_{\omega,x} = \sigma_{x \beta}(\omega) E_{\omega,\beta} \:.
\]
Taking into account that, for the degenerate electron gas with the Fermi energy $\varepsilon_{\rm F}$ and the electrostatic potential $V(x)$ with vanishing average, $\overline{V(x)} = 0$, the local conductivity is a function of the difference $\varepsilon_{\rm F} - V(x)$ we obtain for the dynamic electron density in the first order in $V(x)$
\begin{equation} \label{deltaN}
\delta N_{\omega}(x) = \frac{i}{e \omega} \sum\limits_{\eta = x,y}{\partial \sigma_{x \eta}(\omega) \over \partial \varepsilon_\text{F}} {\partial \over \partial x} \left[V(x)E_{\omega,\eta}(x)\right]\:.
\end{equation}

Hereafter we assume the following inequalities to be satisfied
\begin{equation} \label{inequal}
\varepsilon_\text{F} \tau / \hbar > \omega \tau > \omega_c \tau > 1 > {\rm e}^{-\pi/(\omega_c\tau)}\:,
\end{equation}
where $\omega_c = |e B_z|/(mc)$ is the cyclotron frequency and $m$ is the electron effective mass. As we show in the next section these inequalities correspond to the conditions of the experiments described above. With an accuracy down to $(\hbar\omega_c/\varepsilon_\text{F})\text{e}^{-\pi/(\omega_c\tau)} \ll 1$ the electron density and the Fermi energy are related, in the same way as for zero magnetic field,  by $N = (m/\pi\hbar^2)\varepsilon_\text{F}$. 
Therefore one can rewrite the derivative in Eq.~(\ref{current}) as
\begin{equation} \label{NF}
	{\partial \sigma^0_{\alpha \beta} \over \partial N} \approx \frac{\pi\hbar^2}{m} {\partial \sigma^0_{\alpha \beta}\over \partial \varepsilon_{\rm F}}\:.
\end{equation}
Substituting Eqs.~(\ref{deltaN}) and  (\ref{NF}) into Eq.~(\ref{current}), we arrive at
\begin{align}
\label{DCDRcurrent}
&j_\alpha^{\rm DCDR} =  i{\pi \hbar^2 \over m e \omega}{\partial \sigma^0_{\alpha \beta}\over \partial \varepsilon_\text{F} } {\partial \sigma_{x \eta}(\omega)\over \partial \varepsilon_\text{F}}\\
& \times  \overline{\left\{E^*_{\omega,\beta}(x) {\partial \over \partial x}\left[V(x)E_{\omega,\eta}(x)\right]\right\}} + {\rm c.c.} \nonumber
\end{align}

The near field $E_{\omega,\eta}(x)$ in real superlattices has a complex structure~\cite{ratchet_graphene,PRB2017,Ivchenko_Petrov_FTT}. For  CdTe-based DGG structures used in the above experiments it has been calculated in Ref.~\cite{PRB2017}. Considering a real profile of $E_{\omega,\eta}(x)$ the solution of Eq.~(\ref{DCDRcurrent}) can only be obtained by numerical calculations.  
Moreover, the electrostatic potential $V(x)$ can not only have a more complex modulation but also be comparable with the Fermi energy $\varepsilon_{\rm F}$ giving rise to contributions nonlinear in $V(x)$~\cite{ratchet_graphene}. At the same time, it has been demonstrated in Ref.~\cite{Review_JETP_Lett} that taking the near field and the electrostatic potential as
\begin{align} \label{EVx}
&{\bm E}_{\omega}(x) = {\bm E}_0 \left[ 1 + h_1 \cos{\left( \frac{2 \pi}{d} x + \varphi_E\right)}\right]\:, \nonumber
\\ 
&V(x) = V_1 \cos{\left( \frac{2 \pi}{d} x + \varphi_V\right)}\:.
\end{align}
allows one to avoid cumbersome formulas but describe well the 
main features of the ratchet effects.  Therefore, for transparency of presentation, we confine ourselves in the following to the above expressions to describe the ratchet effect. Then the DCDR current is proportional to the  structure asymmetry parameter $\Xi$, Eq.~\eqref{Xi}, 
\begin{equation}
	\Xi = \overline{{dV \over dx} |{\bm E}_{\omega}(x)|^2} = \frac{2 \pi}{d} V_1 h_1 |{\bm E}_0|^2 \sin{ (\varphi_E -  \varphi_V) }\:.
\end{equation}

In the Shubnikov-de~Haas oscillation regime we have~\cite{Raichev_2008}
\begin{align} \label{SdH}
&\sigma^0_{xx} = {\sigma_0\over \omega_c^2 \tau^2} \left(1 - 4 \delta_c\right),   \sigma^0_{xy} = -  \frac{\sigma_0}{\omega_c \tau} \left( 1   + 
\frac{6  \delta_c}{(\omega_c \tau)^2} \right) \frac{B_z}{|B_z|} \:, \nonumber \\	
& \sigma_{xx}(\omega) = {1+ i\omega\tau \over \omega^2\tau^2}\sigma_0\:,\: 
\sigma_{xy}(\omega)  = {\omega_c\sigma_0 \over \omega^2 \tau} \left(1-{2i\over \omega\tau} \right)\frac{B_z}{|B_z|}\:, 
\end{align}
where $\sigma_0 = e^2N\tau/m$, 
\begin{equation}
	\delta_c = \text{e}^{-\pi/(\omega_c\tau)}  {z\over \sinh{z}}\cos{\left(2\pi\varepsilon_\text{F}\over\hbar\omega_c\right)}\: ,\: z= {2\pi^2k_\text{B}T\over\hbar\omega_c}\:.
\end{equation}
The {\it dc} conductivities $\sigma_{xx}^0, \sigma_{xy}^0$ contain oscillatory contributions proportional to $\delta_c$ which is a function of $\omega_c, \varepsilon_{\rm F}$ and $T$. As compared to these contributions the oscillatory behavior of the \textit{ac} conductivities $\sigma_{xx}(\omega), \sigma_{xy}(\omega)$  is suppressed by an additional factor $\omega_c/\omega \ll 1$. 
Then, accounting for the leading terms only, we obtain the following expressions for the ratchet current
\begin{align}
\label{DCDR}
&	j_x^{\rm DCDR} =  \delta_s \Xi  F\left(2|e_x|^2\omega\tau+ {3\omega\over 2\omega_c}P'_l \frac{B_z}{|B_z|} +\omega_c\tau P_\text{c} \frac{B_z}{|B_z|} \right)\:, \nonumber \\
&	j_y^{\rm DCDR} = \delta_s \Xi  F \left(-{3\omega\over \omega_c}|e_x|^2 \frac{B_z}{|B_z|} + \omega\tau P'_l \right)\:, 
\end{align}
where ${\bm e} = {\bm E}_0/|{\bm E}_0|$ is the polarization unit vector of the electric near field, $P_c$ and $P'_l = e_xe_y^*+e_x^*e_y$ are respectively the degrees of circular polarization and linear polarization in the axes $x',y'$ rotated around $x,y$ by 45$^{\circ}$, 
and
\[
	\delta_s = \text{e}^{-\pi/(\omega_c\tau)}  {z\over \sinh{z}}\sin{\left(2\pi\varepsilon_\text{F}\over\hbar\omega_c\right)}\:,\:
	F = -\frac{4 e^3 \varepsilon_\text{F}}{m (\hbar \omega)^3 \omega_c^3 \tau^2} \:.
\]

\subsection{Radiation-Induced Conductivity Oscillations (RICO)}

While the second mechanism of the magnetic ratchet effect 
relays on a similar physics of the electric current formation 
it implies another origin of the $1/B$-periodic oscillations. The current due to this mechanism  occurs because of
a correction to the conductivity, $\delta \sigma$, linear in the radiation power which is caused by  shifts of the guiding centers of electron cyclotron orbits in real space due to radiation-assisted scattering off disorder similarly to microwave-induced resistance oscillations termed
MIRO~\cite{I_Dmitriev_review}. It is given by~\cite{I_Dmitriev_1st_order}
\begin{equation} \label{RICO}
	\delta\sigma_{\pm} = 6 \delta_c {Ne^2\over m\omega_c^2\tau} \left[{ev_\text{F}|\bm E_{\omega}| \over \hbar\omega\left(\omega\pm \omega_c\frac{B_z}{|B_z|}\right)} \right]^2.
\end{equation}
Here $v_{\rm F} = \sqrt{2 \varepsilon_{\rm F}/m}$ is the Fermi velocity, the upper and lower signs are used for the left- and right-hand circular polarizations, and we replaced $\sin^2{(\pi\omega/\omega_c)}$ by  1/2 because of a large value of the ratio $\omega/\omega_c$. In Ref.~\cite{I_Dmitriev_1st_order}, Eq.~(\ref{RICO}) was used to calculate the magnetophotocurrent in the presence of a \textit{dc} electric field ${\cal E}$. Unlike Refs.~\cite{I_Dmitriev_review,I_Dmitriev_1st_order}, we replace the homogeneous field 
${\cal E}$ by the field $- e^{-1} dV(x)/dx$ of the periodic potential $V(x)$ and take into account the $x$ dependence of the electric near field $\bm E_{\omega}(x)$. Then the RICO contribution to
the ratchet current depending on the circular polarization reads
\begin{equation}
	j^\text{RICO}_{x,P_\text{c}=\pm 1} = -{1\over e} \overline{ \delta \sigma_\pm(x) {dV\over dx}}.
\end{equation}
Under the condition $\omega \gg \omega_c$ relevant to the experiment, for the circular ratchet current defined by ${j^\text{RICO}_{x,\text{circ}}=(	j^\text{RICO}_{x,P_\text{c}=+ 1} -	j^\text{RICO}_{x,P_\text{c}=- 1})/2}$  we obtain
\begin{equation} \label{RICO2}
	j^\text{RICO}_{x,\text{circ}} \approx -\delta_c \Xi F \frac{6}{\pi}\frac{\varepsilon_{\rm F}}{\hbar \omega} \frac{\omega_c^2 \tau}{\omega} P_\text{c} \frac{B_z}{|B_z|}\:.
\end{equation}

\section{Discussion}
\label{discussion}

We begin the discussion with the estimation of relative 
contributions of the two mechanisms described above, DCDR and RICO, considering parameters
relevant to the experiment. In our experimental conditions,  the radiation frequency is fixed at $\omega=1.6 \times 10^{13}$~s$^{-1}$,  the Fermi energy $\varepsilon_\text{F}=15.8$~meV is extracted from the Shubnikov-de~Haas oscillations, the electron scattering time $\tau=0.54$~ps is known from mobility measurements, and the cyclotron frequency at the field strength of $B=2$~T equals to $\omega_c=3.52\times 10^{12}$~s$^{-1}$~\cite{Yakovlev_CR}. With these parameters we obtain that inequalities~\eqref{inequal} used to derive both RICO and DCDR ratchet currents 
are met in the described experiments: $\varepsilon_\text{F}\tau/\hbar \approx 13$, $\omega \tau=8.6$, $\omega_c\tau=1.9$, $\exp[-\pi/(\omega_c\tau)]=0.19$. Comparing two oscillating contributions, described by Eqs.~(\ref{DCDR}) and (\ref{RICO2}), to the circular ratchet current we see that  $j_x^\text{RICO}$ oscillates in phase with Shubnikov--de Haas resistance oscillations while the contribution $j_{x}^\text{DCDR}$ is phase-shifted by $\pi/2$. A ratio of these two contributions is, like an indeterminate form $\infty \times 0$,  
a product of large and small values, see the inequalities~(\ref{inequal}),
\begin{equation}
	{ j^\text{RICO} \over j^\text{DCDR}} \sim {\varepsilon_\text{F}\over \hbar\omega}  \times {\omega_c \over \omega}\:.
\end{equation}
Due to the fact that in the experimental conditions
$\varepsilon_\text{F} \omega_c/(\hbar\omega^2) < 1$ the ratchet current in the studied system
is dominated by the Dynamic Carrier-Density Redistribution mechanism.
This conclusion is consistent with the experimental data presented in Fig.~\ref{Fig_7_bdep_sB}, which demonstrates the phase match between the
magnetic field dependencies of the conductivity derivative, $\partial \sigma/ \partial B$, and the current $J_y$, and therefore the phase shift by $\pi/2$ between the magneto-oscillations in the conductivity and in the polarization dependent ratchet current.

The DCDR current given by Eq.~\eqref{DCDR} describes also other experimental findings. First of all, it is proportional to the lateral asymmetry parameter $\Xi$ which determines the zero-field ratchet current polarity, see Fig.~\ref{Fig_6_alpha_sA}. Particularly, the polarization dependent magnetic ratchet current reverses its direction under the sign inversion of $\Xi$ which is realized by reversing the 
voltage polarities applied to two subgates. Next, the photocurrent $\bm j^\text{DCDR}$ exhibits oscillations periodic in the inverse magnetic field. Moreover, the amplitude of the magneto-oscillations dominates the zero-field ratchet current in accordance with the factor $(\varepsilon_\text{F}/\hbar\omega_c)\exp{(-\pi/\omega_c\tau)}>1$. Both facts are clearly detected in the experiments on linear and circular ratchet effects, see Figs.~\ref{Fig_6_alpha_sA}--\ref{Fig_8_bdep_sB}.

Let us turn now to the polarization dependence of the ratchet current. Equations~\eqref{DCDR} are derived for a system of the C$_s$ symmetry with the mirror reflection plane $xz$, see Appendix~\ref{appendix2}. 
In accordance with the symmetry, the photocurrent contributions $j_x \propto |e_x|^2$ and $j_y \propto P_l'$ are even functions of $B_z$ whereas the contributions $j_x \propto P_l'$ or $P_c$ and $j_y \propto |e_x|^2$ are odd in $B_z$. It is important to note that the unit vector ${\bm e}$ and the polarization degrees $P_l \equiv |e_{x}|^2-|e_{y}|^2, P_l', P_c$ used in the theory are the characteristics of the near field acting on the electrons in the quantum well. 
However, in the experiment the photocurrent components $j_x, j_y$ are measured in dependence on the polarization state of the incident radiation. The degrees of the linear polarization,
\begin{equation}
P_{l0} \equiv |e_{0x}|^2-|e_{0y}|^2\:,\qquad P_{l0}' = e_{0x}e_{0y}^*+e_{0x}^*e_{0y}\:,
\end{equation}
and the circular polarization, $P_{c0}$, are varied as $P_{l0} = (1+\cos{4\varphi})/2$, $P_{l0}' = \sin{4\varphi}/2$, $P_{c0} = \sin{2\varphi}$ when the $\lambda/4$ plate is used, and as $P_{l0} = \cos{2\alpha}$, $P_{l0}' = \sin{2\alpha}$ when the $\lambda/2$ plate is used ($P_{c0}$ is independent of $\alpha$). In passing through the metal grating, the electromagnetic wave undergoes a change in the polarization: due to an effective birefringence,  the circular polarization and the linear polarization in the axes $x',y'$  are partially transformed into each other so that the polarizations $P_{l}' $ and $P_{c}$ become linear combinations of $P_{l0}'$ 
and $P_{c0}$~\cite{Ivchenko_Petrov_FTT}. In particular, the circularly polarized incident radiation is transformed into an elliptical polarization with nonzero value of $P_l'$  and, thus, can induce both the $x$ and $y$ components of the photocurrent.   As a result, both components have contributions proportional to $P_{c0}$, which are detected
 in the experiment, Figs.~\ref{Fig_8_bdep_sB} and \ref{Fig_9_phi_sA}. 
Analyzing the current given by Eq.~(\ref{DCDR}) 
and taking into account the above transformations, we 
obtain that dependencies of the magnetic  ratchet effect on the orientation of the electric field vector of linearly polarized radiation also describe polarization behavior
detected in experiment, Figs.~\ref{Fig_6_alpha_sA} and \ref{Fig_7_bdep_sB}.

Finally we point out that the observed beating  in DMS structures at high magnetic fields and low temperature, see e.g. Fig.~\ref{Fig_6_alpha_sA} are attributed to the 
giant Zeeman splitting, see e.g. Refs.~\cite{Gaj79,Fur88,Dietl,DMS2010,Kossut2011}. The latter, being caused by the exchange interaction with Mn$^{2+}$ ions, affects the $1/B$-periodic oscillations of the magnetoconductivity and, consequently of the magnetic ratchet effect. For polarization independent thermoratchet effect it has been considered in details in Ref.~\cite{PRB2017}. As the beatings have the same origin for Seebeck, linear and circular magnetic ratchet effects they will not be discussed here in more details. 

\section{Summary} \label{summary}

To summarize, we have experimentally demonstrated and
theoretically explained linear and circular magnetic quantum ratchet effects in semiconductor structures with an asymmetric inter-digitated  
gate superlattice on top of the quantum well structures.
Polarization sensitive electric currents driven by terahertz electric field exhibit sign-changing magneto-oscillations with an amplitude giantly enhanced as compared to the photocurrent at zero magnetic field. 
The amplitude and the sign of the ratchet current are controlled by the in-plane orientation of the electric field in respect to the DGG structure (linear magnetic ratchet effect) or by the photon helicity (circular magnetic ratchet effect).  The photocurrent generation mechanism can be
well described in terms of semiclassical theory of magnetic
ratchet effects.  
The theory of the linear and circular magnetic ratchets 
shows that the ratchet currents are caused by the 
Dynamic Carrier-Density Redistribution mechanism and, in agreement with experiment, follow the oscillations of the derivative of the longitudinal conductivity.

\section{Acknowledgments} \label{acknow}

The support from the DFG priority program SFB~689,
the Volkswagen Stiftung Program, and the foundation ``BASIS'' is gratefully acknowledged. 
The work of G.V.B., L.E.G. and E.L.I. is supported by the Russian Science Foundation (project no. 14-12-01067). The research in Poland was
partially supported by the National Science Centre (Poland)
through Grant No. DEC-2012/06/A/ST3/00247 and by the Foundation
for Polish Science through the IRA Programme financed by EU within
SG OP.

\appendix
\section{CIRCULAR EFFECT IN THE SAMPLE \#A}
\label{appendix}

\begin{figure}[ht]
    \includegraphics[width=\linewidth]{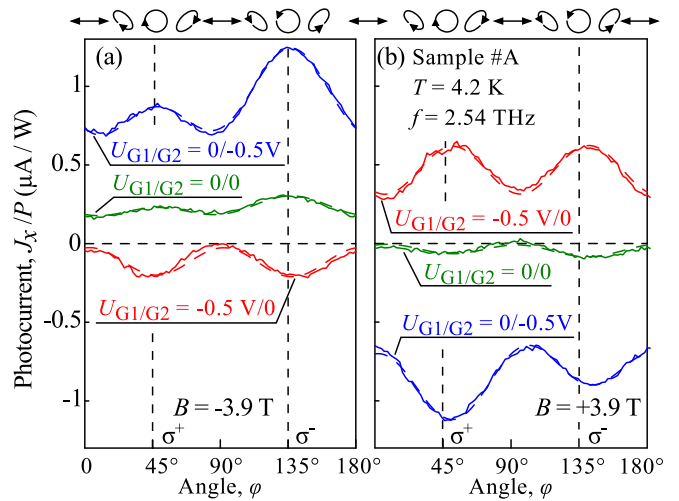}
    \caption{  
Photocurrent as a function of the phase angle $\varphi$ defining the radiation polarization state measured in sample~\#A for two magnetic field strengths corresponding to  the magneto-oscillation peaks. The data are presented for different  asymmetries of the electrostatic potential. The ellipses on top illustrate the polarization states for several values of $\varphi$. Vertical dashed lines indicate angles 
$\varphi$ for right-handed  and left-handed  circularly polarized radiation. Dashed curves are 
the result of fitting the Eq.~(\ref{circ_B0_fit}) to experimental data.
  \label{Fig_9_phi_sA}        }
\end{figure}

Circular magnetic ratchet effect has also been detected in sample~\#A, however, in contrast to the data of sample~\#B it is superimposed with a dominating contribution of the polarization independent Seebeck ratchet current.
Helicity dependence of $J_x,$  shown in Fig.~\ref{Fig_9_phi_sA}
for two values of magnetic field $B = \mp 3.9$~T corresponding to peaks of magneto-oscillations, is  well described by Eq.~(\ref{circ_B0_fit}). While the polarization independent contribution dominates the response
the circular magnetic photocurrent proportional to the lateral degree of asymmetry has also been detected. In Figs.~\ref{Fig_9_phi_sA}(a) and (b) it is most pronounced for the gate voltage combination $U_{{\rm G1}/{\rm G2}} = 0 / \,-0.5$~V.  

\section{SYMMETRY REDUCTION OF THE MODEL STRUCTURE}
\label{appendix2}

Due to the finite length of the stripes, the designed dual grating structure has no $(xz)$ mirror reflection plane and, strictly speaking, its symmetry C$_1$ imposes no restrictions on the polarization dependence of the photocurrent, even at zero magnetic field. 
In this case the polarization dependencies of the ratchet effects 
are given by Eqs.~\eqref{FITalpha} and~\eqref{circ_B0_fit}. 
This behavior has been reported previously  for ratchet effect in similarly designed DGG structures fabricated on top of InAlAs/InGaAs/InAlAs/InP high electron mobility transistors (HEMT)~\cite{Otsuji_Ganichev} and graphene~\cite{ratchet_graphene}. Note that more details on the symmetry analysis of photocurrents in quantum wells of C$_1$ symmetry can 
be found in~Refs.~\cite{Belkov2008,Wittmann}.

\begin{figure}[ht]
        \includegraphics[width=\linewidth]{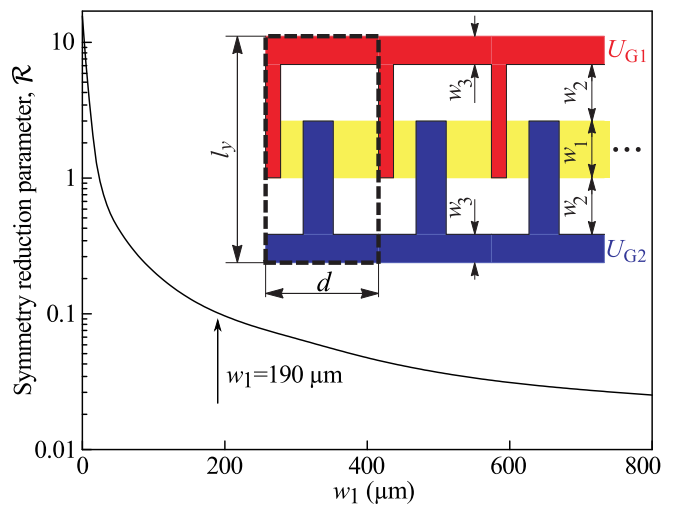}
        \caption{Dependence of the parameter $\cal R$ on the width $w_1$ for fixed  $w_2=190~\mu$m, \textit{w}$_3=30~\mu$m. 
			The inset shows schematics of the periodic lateral superlattice with the supercell enclosed within the dashed rectangle. The voltages $U$ and $-U$ are applied to the narrow  (red) and wide (blue) stripes, the yellow area shows the overlap region where the ratchet structure is realized.}
        \label{Fig_add}
\end{figure}

To estimate the effect of symmetry reduction from C$_s$ to C$_1$ in the DGG structures studied here we consider a model structure shown in the inset in Fig.~\ref{Fig_add}, the supercell is indicated by a dashed rectangle, $w_1, w_2$ and $w_3$ are, respectively, the widths of the ratchet region and the regions where only wide or narrow stripes are present. 
For calculation, we take the structure period $d=14$~$\mu$m, the widths $w_1=w_2= 190$~$\mu$m, and assume the voltages of opposite signs $\pm U$ to be applied to the gates $\rm G1$ and $\rm G2$, the width of both the gates is $w_3=30$~$\mu$m.
The supercell parameters are taken for the sample
\#A metal stripes widths $d_1 = 1.85~\mu$m and  $d_2 = 3.7~\mu$m, first and second spacing between the stripes are   $a_1 = 2.8~\mu$m and $a_2 = 5.65~\mu$m, respectively.

By analogy with the lateral asymmetry parameter $\Xi$, see Eq.~(\ref{Xi}), defined for the stripes of infinite length, we can introduce two parameters
\begin{equation}
\begin{split}
\Xi_x=\dfrac{1}{l_y}\int\limits_0^{l_y} dy\overline{{dV \over dx} |{\bm E}_\omega(\bm{\rho})|^2}\:,\\
\Xi_y=\dfrac{1}{l_y}\int\limits_0^{l_y} dy\overline{{dV \over dy} |{\bm E}_\omega(\bm{\rho})|^2}\:,
\end{split}
\end{equation}
where $\bm{\rho}=(x,y)$ is the in-plane radius-vector, $l_y$  is size of the supercell along the $y$ direction, and the overline still denotes the average over $x$. The parameters $\Xi_x$ and $\Xi_y$ describe the systems's asymmetry along the $x$  and $y$ directions, respectively. In the limit $w_1 \to \infty$ with $w_2, w_3$ being fixed, the parameter $\Xi_x$ tends to $\Xi$ and the parameter $\Xi_y$ vanishes. Evidently, the degree of symmetry reduction from C$_s$ to C$_1$ can be estimated by the ratio ${\cal R}=\Xi_y/\Xi_x$.

In order to calculate the distribution of electric near field in the QW plane we follow Refs.~\cite{Popov2010,Ivchenko_Petrov_FTT} and apply the approximate method of surface currents. 
We generalize this method by considering the transmitted and reflected electromagnetic waves as functions of not one but two in-plane coordinates $x$ and $y$. In this approach, the height of the stripes $h$ 
is considered small enough to treat them as infinitely thin and use the following boundary conditions for the radiation magnetic field above and below the grating
\begin{align}
\begin{split}
&B_x(\bm{\rho},+0)-B_x(\bm{\rho},-0)=4 \pi j_y(\bm{\rho})/c\:, \\
&B_y(\bm{\rho},+0)-B_y(\bm{\rho},-0)=-4 \pi j_x(\bm{\rho})/c\:.
\end{split}
\end{align}
Here the density of {\it ac} two-dimensional current is given by
$\bm{j}=\sigma(\bm{\rho}) h \bm{E}(\bm{\rho})$, where, at the metallic stripes, $\sigma(\bm{\rho})$ is the bulk conductivity of metal (gold)  and, outside the stripes, $\sigma(\bm{\rho})=0$.

For the calculations of the electrostatic potential $V(\bm{\rho})$ we use a reasonable approximation  of two-dimensional Laplace's equation
$\partial^2 V/\partial x^2+\partial^2 V/\partial y^2 =0$ with the boundary conditions: $V(\bm{\rho})=-U$ at the edge of wide gate stripes (shown in blue in the inset in Fig.~\ref{Fig_add})
and $V(\bm{\rho})=+U$ at the edge of narrow stripes (red).

The functions $|{\bm E}_{\omega}(\bm{\rho})|^2$ and $V(\bm{\rho})$ are found numerically. As one could expect, the main contribution to the parameter $\Xi_x$ comes from the yellow area in the inset in Fig.~\ref{Fig_add} while only the regions lying close to the end of the stripes contribute to $\Xi_y$. 
With the increasing $w_1$, the parameter ${\cal R}$ goes to zero.
The calculation shows that the parameter ${\cal R}$ does not exceed 0.1 and the point-group symmetry of the model structure shown in the inset in Fig.~\ref{Fig_add} is likely to be $C_s$ rather than $C_1$ and using this symmetry for the theory presented in Sec.~\ref{theory} is validated. \newline
Figure~\ref{Fig_add} shows the dependence of the symmetry reduction parameter
 $\cal R$ on the width $w_1$ for fixed $w_2=190~\mu$m, $w_3=30~\mu$m.
 With the increasing $w_1$ the parameter $\cal R$ rapidly goes to zero.
 For the experimentally studied structure $w_1=190~\mu$m, ${\cal R}
 \approx 0.1$. This demonstrates that the point-group symmetry of the
 model structure of the inset in Fig.\,\ref{Fig_add} is likely to be $C_s$ rather than
 $C_1$. This validates using this symmetry for the theory developed in
 Sec.\,\ref{theory}.

\FloatBarrier


\begin{thebibliography}{99}

\bibitem{prost97} F. J\"{u}licher, A. Ajdari, and J. Prost, Rev. Mod. Phys. {\bf 69}, 1269 (1997).

\bibitem{reimann} P. Reimann,Phys. Rep. {\bf 361}, 57 (2002).

\bibitem{hanggi} P. H\"{a}nggi and F. Marchesoni, Rev. Mod. Phys. {\bf 81}, 387 (2009).

\bibitem{Denisovhanggi}S. Denisov, S. Flach, and P. H\"{a}nggi, Phys. Rep. {\bf 538}, 77 (2014).

\bibitem{buttiker} M. B\"uttiker, Z. Phys. B {\bf 68}, 161 (1987).
%
\bibitem{buttiker2} Ya.~M. Blanter and M. B\"uttiker, Phys. Rev. Lett. {\bf 81}, 4040 (1998).
%
\bibitem{grifoni} P. Reimann, M. Grifoni, and P. H\"anggi, Phys. Rev. Lett. {\bf 79}, 10 (1997).
%
\bibitem{kotthaus} A. Lorke \emph{et al}., Physica B {\bf 249}, 312 (1998).
%
\bibitem{samuelson} A.~M. Song ~\emph{et al}., Appl. Phys. Lett. {\bf 79}, 1357 (2001).
%
\bibitem{Chepel} A. D. Chepelianskii, M. V. Entin, L. I. Magarill, and D. L.
Shepelyansky, Eur. Phys. J. \textbf{56}, 323 (2007).
%
\bibitem{Chepelianskii}  A.~D. Chepelianskii, M.~V. Entin, L.~I. Magarill, and D.~L. Chepelyansky, Physica E (Amsterdam) {\bf 40}, 1264 (2008).
%
\bibitem{Kvon} S. Sassine, Yu. Krupko, J.-C. Portal, Z. D. Kvon, R. Murali, K. P. Martin, G. Hill, and A. D. Wieck, Phys. Rev. B \textbf{78}, 045431 (2008).
%
\bibitem{Olbrich_PRL_09}

P. Olbrich, E. L. Ivchenko, R. Ravash, T. Feil, S. Danilov,
J. Allerdings, D. Weiss, D. Schuh, W. Wegscheider, and S. D. Ganichev,
Phys. Rev. Lett. \textbf{103}, 090603 (2009).
%
\bibitem{Popov2010} D.V. Fateev, V.V. Popov, and M.S. Shur, Semiconductors {\bf 44}, 1406 (2010).
%
\bibitem{Olbrich_PRB_11} 
P. Olbrich, J. Karch, E. L. Ivchenko, J. Kamann, B. M\"{a}rz, M. Fehrenbacher, D. Weiss, and S. D. Ganichev,
Phys. Rev. B \textbf{83}, 165320 (2011).
%
\bibitem{Review_JETP_Lett} E. L. Ivchenko and S. D. Ganichev,
{Pisma v ZhETF} \textbf{93}, 752 (2011) [{JETP Lett.} \textbf{93}, 673 (2011)].
%
\bibitem{KannanPortal12} E. S. Kannan, I. Bisotto, J.-C. Portal, T. J. Beck, and L. Jalabert, Appl.
Phys. Lett. \textbf{101}, 143504 (2012).
%
\bibitem{Nalitov}  A. V. Nalitov, L. E. Golub, and E. L. Ivchenko, Phys. Rev. B \textbf{86}, 115301 (2012).
%
\bibitem{PopovAPL2013} V. V. Popov, Appl. Phys. Lett. \textbf{102}, 253504 (2013).

\bibitem{Rozhansky2015} I. V. Rozhansky, V. Yu. Kachorovskii, and M. S. Shur,
Phys. Rev. Lett. \textbf{114}, 246601 (2015).
%
\bibitem{PopovIvchenko} V. V. Popov, D. V. Fateev, E. L. Ivchenko, and S. D. Ganichev,
Phys. Rev. B \textbf{91}, 235436 (2015).
%
\bibitem{ratchet_graphene} P. Olbrich, J. Kamann, M. K\"{o}nig, J. Munzert, L. Tutsch, J. Eroms, D. Weiss,
M.-H. Liu, L. E. Golub, E. L. Ivchenko, V. V. Popov, D. V. Fateev, K. V. Mashinsky, F. Fromm, Th. Seyller, and S. D. Ganichev, 
Phys. Rev. B \textbf{93}, 075422 (2016).
%
\bibitem{PRB2017} P.\,Faltermeier,  G.V. Budkin, J.\,Unverzagt, S.\,Hubmann, A.\,Pfaller,
V.V.\,Bel'kov,  L.E. Golub, E.L. Ivchenko,  Z. Adamus, G. Karczewski, T. Wojtowicz,
V.V.\,Popov, D.V.\,Fateev, D.A. Kozlov,
D.\,Weiss, and S.D.\,Ganichev,
Phys. Rev. B \textbf{95}, 155442 (2017).

%
\bibitem{Popov_Otsuji_Knap} V. V. Popov, D. V. Fateev, T. Otsuji, Y. M. Meziani, D. Coquillat,
and W. Knap, Appl. Phys. Lett. \textbf{99}, 243504 (2011).

\bibitem{otsuji} T. Watanabe, S. A. Boubanga-Tombet, Y. Tanimoto, D. Fateev, V. Popov, D. Coquillat, W. Knap, Y. M. Meziani, Yuye Wang, H. Minamide, H. Ito, and T. Otsuji,
IEEE Sensors J. {\bf 3}, 89 (2013). 
%
\bibitem{det2} Y. Kurita, G. Ducournau, D. Coquillat, A. Satou, K. Kobayashi,
S. Boubanga Tombet, Y. M. Meziani, V. V. Popov, W. Knap,
T. Suemitsu, and T. Otsuji, Appl. Phys. Lett. \textbf{104}, 251114
(2014).
%
\bibitem{otsuji2} S. A. Boubanga-Tombet, Y. Tanimoto, A. Satou, T. Suemitsu, Y. Wang, H. Minamide, H. Ito,
D. V. Fateev, V. V. Popov, and T. Otsuji,
Appl. Phys. Lett. {\bf 104}, 262104 (2014).

\bibitem{ellipticitydetector}
S. N. Danilov, B. Wittmann, P. Olbrich, W. Eder, W. Prettl, L. E.
Golub, E.V. Beregulin, Z. D. Kvon, N. N. Mikhailov, S. A.
Dvoretsky, V. A. Shalygin, N. Q. Vinh, A. F. G. van der Meer, B.
Murdin, and S. D. Ganichev, J. Appl. Phys. \textbf{105}, 013106
(2009).
%
\bibitem{ellipticitydetector2}
S. Dvoretsky, N. Mikhailov, Y. Sidorov, V. Shvets, S. Danilov, B.
Wittman, and S. Ganichev,
J. Electron. Mat. \textbf{39},  918 (2010).
%


\bibitem{ellipticitydetector3}
C.~Drexler,  N. Dyakonova, P. Olbrich, J.~Karch, M.~Schafberger, K.~Karpierz, Yu.~Mityagin, M.~B.~Lifshits, 
F.~Teppe, O.~Klimenko, Y.~M. Meziani, W.~Knap, and S.~D.~Ganichev,
J. Appl. Phys.  \textbf{111},  124504 (2012).

\bibitem{Otsuji_Ganichev} P. Faltermeier, P. Olbrich, W. Probst, L. Schell, T. Watanabe,
S. A. Boubanga-Tombet, T. Otsuji, and S. D. Ganichev,
J. Appl. Phys. \textbf{118}, 084301 (2015).
%



\bibitem{JETP_Lett_review} G. V. Budkin, L. E. Golub, E. L. Ivchenko, and S. D. Ganichev, JETP Lett. \textbf{104}, 649 (2016).

\bibitem{Budkin_Golub} 
G.V. Budkin and L.E. Golub, Phys. Rev. B \textbf{90}, 125316 (2014).


\bibitem{Crooker}  S. A.~Crooker,
D.~A.~Tulchinsky, J.~Levy, D.~D.~Awschalom, R.~Garcia, and
N.~Samarth, Phys. Rev. Lett. \textbf{75}, 505 (1995).

\bibitem{Egues} J. C.~Egues and J. W.~Wilkins, Phys. Rev. \textbf{58}, R16012 (1998).

\bibitem{Jaroszynski2002} J. Jaroszynski,
T. Andrearczyk, G.~Karczewski, J.~Wr\'{o}bel, T.~Wojtowicz,
E.~Papis, E.~Kaminska, A.~Piotrowska, D.~Popovic, and T.~Dietl,
Phys. Rev. Lett. \textbf{89}, 266802 (2002).

\bibitem{DMSPRL09} 
S. D.~Ganichev, S. A. Tarasenko, V. V. Bel'kov, P. Olbrich, W. Eder, D. R.~Yakovlev, V. Kolkovsky, W. Zaleszczyk,
G.~Karczewski, T. Wojtowicz, and D. Weiss,
Phys. Rev. Lett. \textbf{102}, 156602 (2009).


\bibitem{Kneip2006} M. K. Kneip, D.~R.~Yakovlev,  M.~Bayer, G.~Karczewski, T.~Wojtowicz, and J.~Kossut,
Appl. Phys. Lett. \textbf{88}, 152105 (2006).


\bibitem{Gaj79} J. A. Gaj, R. Planel, and
G. Fishman, Solid State Commun. \textbf{29}, 435 (1979).

\bibitem{Fur88} J. K. Furdyna, J. Appl. Phys. \textbf{64}, R29 (1988).

\bibitem{staab2015} M. Staab, M. Matuschek, P. Pereyra, M. Utz, D. Schuh, D. Bougeard,
R. R. Gerhardts, and D. Weiss,
New J. Phys. \textbf{17}, 043035 (2015).

	\bibitem{Ganichev93} 
	S.D. Ganichev, W. Prettl, P.G. Huggard,
Phys. Rev. Lett. \textbf{71},  
3882
(1993).


\bibitem{Kvon2008} 
Z.\,D. Kvon, S.\,N. Danilov, N.\,N. Mikhailov, S.\,A. Dvoretsky, and S.\,D. Ganichev,
Physica E \textbf{40}, 1885 (2008).

\bibitem{Kohda2012} M. Kohda, V. Lechner, Y. Kunihashi, T. Dollinger, P. Olbrich, C. Sch\"{o}nhuber, 
I.~Caspers, V.V.~Bel'kov, L.E. Golub, D. Weiss, K. Richter, J. Nittaand S.D. Ganichev
Phys. Rev. B Rapid Communic. \textbf{86}, 081306 (2012).




\bibitem{Ganichev1999} 
S.D. Ganichev,
 	Physica B \textbf{273-274},  737 (1999).
	





\bibitem{Schneider04} 
P. Schneider, J. Kainz, S.D. Ganichev, 
V.V.~Bel'kov,
S.N.~Danilov, M.M.~Glazov, L.E.~Golub, U.~R\"{o}ssler,
W.~Wegscheider, D.~Weiss, D.~Schuh, and W.~Prettl, 
J. Appl. Phys.  {\bf 96}, 420 (2004).


\bibitem{book} S.~D. Ganichev, W. Prettl, \textit{Intense Terahertz Excitation of Semiconductors}
(Oxford University Press, Oxford, 2006).



\bibitem{Stokes1}  B.E.A.~Saleh and M.C.~Teich, \textit{Fundamentals of Photonics} (John Wiley \& Sons, Inc., 2007).


\bibitem{BelkovJPCM} V.\,V.~Bel'kov, S.\,D.~Ganichev, E.\,L.~Ivchenko, S.\,A.~Tarasenko,
W.~Weber, S.~Giglberger, M.~Olteanu, H.-P.~Tranitz,
S.\,N.~Danilov, Petra~Schneider, W.~Wegscheider, D.~Weiss, and
W.~Prettl, J. Phys.: Condens. Matter \textbf{17}, 3405 (2005).



\bibitem{Karch_PRL_2011}    J. Karch, C. Drexler, P. Olbrich, M. Fehrenbacher, M. Hirmer, M. M. Glazov,
S. A. Tarasenko, E. L. Ivchenko, B. Birkner, J. Eroms, D. Weiss, R. Yakimova, S. Lara-Avila, S. Kubatkin,
M. Ostler, T. Seyller, and S. D. Ganichev,
Phys. Rev. Lett. \textbf{107}, 276601 (2011).

\bibitem{Glazov2014}  M. M.\,Glazov and S. D.\,Ganichev,
Phys. Rep. \textbf{535},  101 (2014).


\bibitem{footnotetau} The same assumption has been made to derive Eq.~(\ref{current2}) for ratchet effects at zero magnetic field,
see Eqs.~(12) and (22) in Ref.~\cite{Review_JETP_Lett}.


\bibitem{Raichev_2008} O. E. Raichev, Phys. Rev. B \textbf{78}, 125304 (2008).

\bibitem{Ivchenko_Petrov_FTT} E. L. Ivchenko and M. I. Petrov, Phys. Solid State \textbf{56}, 1833
(2014).

\bibitem{I_Dmitriev_review} I. A. Dmitriev, A. D. Mirlin, D. G. Polyakov, and M. A. Zudov,
Rev. Mod. Phys. \textbf{84}, 1709 (2012).

\bibitem{I_Dmitriev_1st_order} I. A. Dmitriev, 
J. Phys. Conf. Ser. \textbf{334}, 012015 (2011).


\bibitem{Yakovlev_CR}  A. A. Dremin, D. R. Yakovlev, A. A. Sirenko, S. I. Gubarev, O. P. Shabelsky, A. Waag, and M. Bayer, Phys. Rev. B
 \textbf{72}, 195337 (2005).








\bibitem{Dietl} T.~Dietl, in \textit{Handbook on Semiconductors}, vol.\,3b,
Ed. T.S. Moss (North-Holland, Amsterdam, 1994).

\bibitem{DMS2010} \textit{Introduction to the Physics of Diluted Magnetic Semiconductors},
Eds. J. Kossut and J.\,A. Gaj  (Springer, Berlin 2010).

\bibitem{Kossut2011}  \textit{Introduction to the Physics of Diluted Magnetic Semiconductors}, eds. J.~Kossut and J.A. Gaj  (Springer 2011).

%

\bibitem{Belkov2008}
V.\,V. Bel'kov and S.\,D. Ganichev,
Semicond. Sci. Technol. \textbf{23}, 114003 (2008).

\bibitem{Wittmann}
B.~Wittmann, S.N.~Danilov,  V.V.~Bel'kov, S.A.~Tarasenko, E.G.~Novik, H.~Buhmann, C.~Br\"{u}ne, L.W.~Molenkamp, E.L.~Ivchenko, Z.D. Kvon, N.N. Mikhailov, S.A. Dvoretsky, N.\,Q.\,Vinh, A.\,F.\,G.~van~der~Meer, B.~Murdin, and S.D.~Ganichev
Semicond. Sci. and Technol. \textbf{25}, 095005 (2010).



%



%















\end{thebibliography}
\end{document}